\documentstyle[12pt]{article}
\begin{document}
\baselineskip .8cm

\newcommand{\be}{\begin{equation}}
\newcommand{\ee}{\end{equation}}
\newcommand{\eps}{\varepsilon}
\newcommand{\z}{\zeta}
\newcommand{\cc}{\xi}

\title{Ground state wave function  of the  Schr\"odinger equation  in a time
periodic potential.} 
\author{Stefano Galluccio and  Yi-Cheng Zhang \\Institut de Physique 
Th\'eorique, Universit\'e de Fribourg,\\ CH -- 1700,  Switzerland}
\date{}
\maketitle

\abstract{ 
Using a generalized transfer matrix method we exactly solve 
the  Schr\"odinger equation in a   time periodic potential,
with  discretized  Euclidean space-time. The ground state
wave function  propagates in space and time with an oscillating 
soliton-like wave packet and the wave front is wedge shaped. In a
statistical mechanics framework our solution represents the 
partition sum of a directed polymer  subjected to a  potential
layer with alternating (attractive and repulsive) pinning centers.
}

\vspace{1cm}
PACS numbers: 03.65.Ge; 68.35.Rh
\newpage

Schr\"odinger equation   plays a central role in modern physics. In the 
Euclidean space-time,  the time is imaginary and the equation has 
interpretation in statistical mechanics. In contrast  to the stationary case,
i.e.  equations in time independent  potentials,   where 
many  exact solutions exist, for the general  {\it time-dependent}  situation
exact results have been lacking so far.

In this work we present an exact solution, albeit in the Euclidean 
lattice space-time.  We consider a periodic potential alternating between 
attractive and repulsive, placed at the space origin $x=0$. The quantum
analogue would be that of a particle in a forced potential with energy 
pumped  in and out periodically.
This problem has several 
links with important branches of modern theoretical physics.
In the statistical mechanics framework our system, with a delta-like 
potential, is  
usually employed in the lattice  models  for  wetting and depinning
transitions of  directed polymers \cite{fln91},
 as well as it represents a simplified version
 (but nevertheless non trivial) of the  KPZ  equation  for 
kinetic interface growth \cite{kpz}.
In a pure QM context the present problem  has applications in: 
wave function collapse in mesoscopic systems,
 Mott-type hopping, transport in disordered systems, quantum electronics, 
time-dependent Zener tunneling (see \cite{azbel} and references therein).
Moreover  it has been recently pointed out  that  the Schr\"odinger
and diffusion equations are very deeply linked if represented on a lattice.
In fact one can prove that, apart  from the usual analytic  continuation,
they represent two aspects of the {\it same} probabilistic problem 
on a lattice manifold \cite{ord}. This maybe provides  a new interesting 
interpretation of  non-relativistic QM on a lattice.

The strategy of our approach is the following: we assume that at large 
times the wave function has the same periodicity of the 
potential.  We  also require that  it is normalized for all   times.
For  the symmetric  case, i.e.  when  attractive and repulsive potentials
have equal strength, the wave function is localized
around the origin \cite{nz}. The normalized solution oscillates  in the time 
direction and its amplitude  is weaker and weaker  when 
the spatial distance from the origin is larger and larger.
Our wave function also shows that the wave front  has a wedge
shape as would be required   by causality: the potential 
influence can only travel with finite velocity. 

Let us start by  considering the imaginary-time  $d$-dimensional
Schr\"odinger equation
\be
\partial_t \Psi(x,t)=\left(\nabla^2 +\delta(x)\cos(\omega t)\right)
\Psi(x,t),
\ee
with the normalization $\int |\Psi(x,t)|^2\, dx=1$.
Here we have considered  the symmetric case: the time average of
the potential is zero. However our general solution  also includes 
 the nonsymmetrical  situation.
For sake of simplicity we will use the 1-$d$ notation for variables 
and operators even if we  can work in a high dimensional space. 

It can be shown that the above equation can be derived from a Hamiltonian
\cite{nz},\cite{gg95} which describes an elastic chain, 
or a directed polymer,
in a periodic potential (alternatively attractive and repulsive). 
The Hamiltonian reads
\be
 {\cal H}_L(\{h_k\})=J\sum_{k=1}^L |h_{k+1}-h_k|^p-\sum_{k=1}^L 
u_k\delta_{h_k,0},
\ee
and, more specifically,  it gives the energy of a polymer in a 
$(d+1)$-dimensional space 
under the competing effect of the pure tension term proportional to
$J$ and the potential layer at  the origin.
Here $h_k$ defines the position  of the polymer at ``time'' $k$, while  
 $u_k$ is  set to  $u>0$ for even $k$ and $-v<0$ for odd $k$.
 In the usual lattice  version,  one 
introduces RSOS conditions, that is the  height difference $ |h_{k+1}-h_k|$  can
only take values 0 or 1 and overhangs  are forbidden. 
Mathematically  speaking, this is completely equivalent to  assume 
$p=\infty$ \cite{fln91}.
The first term in (1) discourages large humps
of the polymer (it corresponds to the laplacian in (1)),
while the second term   gives 
a positive (resp. negative) contribution to the total energy when  the  
particle passes through the origin for odd (resp. even) times $k$.
Hence  it is the discretized (and non symmetric) form of 
the continuous sinusoidal 
potential  in the original Schr\"odinger equation.

The equilibrium state of the polymer,  whose energy is defined in (2) 
for a given
configuration,  is  the result of the competition between two contrasting 
effects. At temperatures $T>0$ we know from general arguments
\cite{fln91} that large deviations from linearity 
increase the configurational
entropy, but these jumps have also the effect of increasing the internal 
energy  and a phase transition occurs at a given $T_c$
when the cost in energy is
exactly balanced by the gain in entropy. This mechanism is 
completely specified by the free energy (per unit length) $f=F/L$.  
From a geometrical point of view,  below $T_c$,  the polymer is 
{\it localized}
in the sense that its mean variance from the origin remains finite
in the thermodynamic limit $L\rightarrow \infty$. 
The opposite stands at $T>T_c$  in which
the  directed line behaves as a pure random walk in a $d$-dimensional 
space and performs thermal wandering.
The QM counterpart of this phase
transition is the binding/unbinding transition between  eigenstates
of the Hamiltonian defined in (1). 
After having clarified  the mapping, our goal consists in the search
of the partition function:
\be 
{\cal Z}_L(x)=\sum_{\{h_k\}}e^{-{\cal H}_L(\{h_k\})/T},
\ee
which is the discretized  Feynman path integral representation
of the wave function associated to (1).

We  then start by  introducing  the evolution equation for ${\cal Z}_L(x)$
 by means of a  transfer matrix approach:
\begin{eqnarray}
{\cal Z}_{L+1}(x)&=&\left[1+(a_{L+1}-1)\delta_{ x,0}
\right]\\
& \times & 
\left[{\cal Z}_L( x)+t\sum_{i=1}^L\left(
{\cal Z}_L(x+1)+
{\cal Z}_L(x-1) \right) \right]. \nonumber
\end{eqnarray}
The search of the ground state of  eq.(1)  is equivalent to the
determination of  the stationary solution of this recursion equation for 
$L\rightarrow\infty$.
In the thermodynamic limit we know that only the largest eigenvalue  $\eps$
of the transfer matrix (4) gives a significant contribution to the 
free energy density, that is  we have $f\simeq -\log\eps$ .
We recall that, due to the particular form of the
potential layer, we expect to find two distinct ground state
wave functions in the bound regime, one for even ($\Psi_+$) and one 
for odd ($\Psi_-$) times. 

The first step consists in the search of a self-consistent expression 
for the partition ``wave'' function. 
The full calculation has been already  carried out 
in a general hypercubic lattice \cite{gg95}: the result is that  for any set
of free parameters $\{d,u,v,J\}$  the normalized partition functions in the
Fourier space  $G_L^\pm (k)={\cal Z}_L^\pm(k)/(1+2dt)^L$  
(with $t=\exp(-J/T)$)  \cite{foot1}  satisfies 
the following implicit equation
\be 
G_{L+1}^\pm(k)=G_L^\mp \xi(k)+{\cal A}\int_0^1d^d\, q\,  G_L^\mp(q)\xi(q),
\ee
where ${\cal A}=\exp(u/T)-1=A^+$ for  even  times and 
${\cal A}=\exp(-v/T)-1=A^-$ for odd times.  Moreover we have introduced the 
quantity $\xi(k)=[1+2t\sum_{i=1}^d \cos(\pi k_i)]/(1+2dt)$.
To take into account the  periodicity of the potential layer we 
 apply transformation (5)  twice,  so that we can  relate 
$G_{2L+2}^+$ to $G_{2L}^+$ and  $G_{2L+1}^-$ to $G_{2L-1}^-$. 
By using the ansatz,
 valid  at large $L$, 
$G_{L}^\pm(k)=\eps^2 G_{L-2}^\pm(k)$ we finally get \cite{gg95}
\begin{eqnarray}
G_+(k) &= &\frac{A^+}{\eps^2-\xi^2}K_2^++
\frac{A^+A^-}{(1+2dt)(\eps^2-\xi^2)}K_1^+ \nonumber \\
& + &\frac{A^-\xi}{\eps^2-\xi^2}K_1^+.
\end{eqnarray}
where we have introduced the two constants
\be
K_n^+=\int_0^1d^d\, q \; \xi^n(q) G_+(q), \quad n=1,2.
\ee
We  simply find, by symmetry,  that $G_-(k)$  can be derived from 
$G_+(k)$ by 
interchange between  $A^+$ and $A^-$.
A phase transition occurs when the maximum eigenvalue converges 
towards $1^+$ \cite{nz}, and  it can be studied by means of the 
{\it mass gap} $\mu$  defined as the inverse of the transversal 
correlation length, i.e.   $\Psi(x)\simeq \exp(-\mu x)$. One can prove that
$\eps \simeq 1+\mu^2$ near the transition point \cite{nz}.
The r.h.s. of eq.(6)  implicitly depends on  $G^\pm(k)$, 
 the full knowledge of the wave functions would  therefore require
4 independent equations for the constants $K_{1,2}^\pm$. The
first couple of equations can be simply derived from (6) by multiplying
both sides for $\xi(k)$ and integrating over the $k$-momentum.
At the end we  obtain   the following identity 
\be
K_1^\pm\left(A^\mp{\cal I}_2 +\frac{A^\pm A^\mp}{1+2dt} {\cal I}_1-1
\right)+K_2^\pm A^\pm {\cal I}_1=0.
\ee
where 
\begin{eqnarray}
{\cal I}_1 &= &\frac{1+2dt}{4t}(f_1(0)-g_1(0)),\nonumber \\
{\cal I}_2 &= &\frac{1+2dt}{4t}(f_1(0)+g_1(0))-1,
\end{eqnarray}
and
\begin{eqnarray}
f_\eps(x) &= &\int_0^\infty du e^{-\eps_1 u}\, I_{|x|}(u)^d, \quad \nonumber \\
g_\eps(x) &= &\int_0^\infty du e^{-\eps_2 u}\, I_{|x|}(-u)^d.
\end{eqnarray}
In the above   we have introduced  the two quantities
 $\eps_1=[\eps(1+2dt)-1]/2t$, $\eps_2=[\eps(1+2dt)+1]/2t$ and 
we have used an integral representation of the modified Bessel
function of integer order $I_n(u)=\int_0^1 \exp\{u \cos y\} \cos{ny}\, dy$.
It is  simple to  convince oneself
that any other attempt to find close equations for these constants
from (6) would lead to relations which are not linearly independent
respect to (8). 

For the meanwhile we neglect the  search of the second couple 
of equations  and we concentrate on equation (6), which must be 
antitransformed in order to get  the real space form  
of our solution.
In performing  the calculation we encounter  non-trivial integrals of the 
form:
\be 
{\cal J}_n(x)=\int_0^1d^d\, k \prod_{i=1}^d \cos(\pi k_ix_i)\frac{\xi(k)^n}
{\eps^2-\xi^2(k)}, \quad n=0,1.
\ee
Without entering into mathematical details, we simply note that,
by  means of an appropriate Feynman  integral representation,
we can simplify  them and after  some more algebra we finally 
obtain that 
\begin{eqnarray}
{\cal J}_0(x) &= &\frac{1+2dt}{4\eps t}\left[f_\eps(x)+g_\eps(x)\right], 
\nonumber \\
{\cal J}_1(x) &= &\frac{1+2dt}{4t}\left[f_\eps(x)-g_\eps(x)\right].
\end{eqnarray}
The symmetry of the system tells us that any  function 
of $x$ must be invariant under axis reflection; this is why
in the above integrals only the absolute value of $x$ appears.  
Note that after this manipulation, the dependence
of our wave function on the spatial variable $x$ comes  only 
from the order of the Bessel function involved in the integrals.
Moreover, by using the property that $I_{|x|}(u)$ is an even (odd) function
of the argument for even (odd) values of $|x|$ it is clear that 
our ground state solution has a particular  oscillation as a function
of $|x|$. As we will discuss below, this is  nothing but a direct 
consequence of the alternating potential  we have put at the origin.
We also note that $f_\eps(x)$ is a well known integral in the statistical
mechanics context \cite{yd}: if one  takes $\eps=1$ 
(or $\eps_1=d$),  it is associated  to the total probability that 
a random walker, 
started from the origin, could   finally reach a point  $x$ in 
a $d$-dimensional
 cubic lattice. The  divergence  of $f_1(x)$   for $d<3$  has also important 
consequences in the polymer  depinning  framework
\cite{gg95}.

To summarize,  our non-normalized wave function  reads
\be
\Psi_\pm(x)=\left(A^\pm K_2^\pm+\frac{A^\pm A^\mp}
{1+2dt}K_1^\pm\right){\cal J}_0(x)+
A^\mp K_1^\pm{\cal J}_1(x).
 \ee
The above formula gives the qualitative behavior of our solution,  but  
to get  the full  normalized wave function we also need   a 
closed form for $K_{1,2}^\pm$.
As  we know that  $\eps>1$  for a non zero mass gap,  we can  restrict
to the localized phase and  impose the correct normalization of  the 
wave function.  Therefore we ask that $\sum_{\{x\}} \Psi_\pm(x)^2=1$ in the
real space, or equivalently, $\int_0^1 d^dk\, G_\pm(k)^2=2$ in the 
momentum space. This condition would give us the second couple of equations
we needed to  find    $K_{1,2}^\pm$.
Unfortunately  the integrals resulting  from  (6) after this manipulation
are much harder to handle with. In fact we find that the normalization
condition reads
\begin{eqnarray}
F_2(K_1^\pm, A^\mp)M_2+F_1(K_{1,2}^\pm, A^\pm , A^\mp , d, t)M_1 
&+& \nonumber \\
F_0(K_{1,2}^\pm ,A^\pm , A^\mp ,d,t)M_0 & = & 2,
\end{eqnarray}
 where $F_0,F_1,F_2$ are some algebraic functions of their arguments
and 
\be
M_n=\int_0^1d^dk \, \frac{\xi^n(k)}{\left[\eps^2-\xi^2(k)\right]^2},
\quad n=0,1,2.
\ee  
We  believe that no simple way can be found to simplify these
high-dimensional integrals, and then at this point we are forced to 
restrict ourselves to the 1-$d$ case. In a pure statistical mechanics
framework one could, in principle, ask 
 $\Psi_\pm(x)$ to be sommable and not square-sommable. In this case
we can perform all integrations and  get  the {\it complete} solution in all
 dimensions. 
Nevertheless, in the spirit of QM,  in the present study we 
prefer  to  use the 
usual normalization  of the  square of  $\Psi_\pm(x)$ .  
At $d$=1 all previous integrals can be exactly solved and we find that
\begin{eqnarray}
f_\eps(x) &=&\frac{1}{\sqrt{\eps_1^2-1}\left(\eps_1+
\sqrt{\eps_1^2-1}\right)^{|x|}},
\nonumber \\
g_\eps(x) &=& \frac{(-1)^{|x|}}
{\sqrt{\eps_2^2-1}\left(\eps_2+\sqrt{\eps_2^2-1}\right)^{|x|}}.
\end{eqnarray}
From these formulas we get immediately the result that the mass gap in the 
localized phase is given by $\mu=\log(\eps_1+\sqrt{\eps_1^2-1})$ 
and then it vanishes, near the transition as $\sqrt{\eps-1}$. 
This is, {\it a posteriori}, the proof that the limit $\eps\rightarrow 1^+$
gives the transition point, as previously stated.

By using (8) and (14) we can finally solve for $K_{1,2}^\pm$:
\begin{eqnarray}
K_1^\pm & = & \sqrt{2}{\cal I}_1\left\{
\left[1-2A^\mp {\cal I}_2+(A^\mp)^2{\cal I}_2^2\right]M_0\right. \nonumber\\
& + & \left. \left[2A^\mp{\cal I}_1-2(A^\mp)^2{\cal I}_1{\cal I}_2\right]M_1
+(A^\mp)^2{\cal I}_1^2M_2\right\}^{-1/2}, \nonumber  \\
K_2^\pm &= &\left(\frac{1}{A^\pm{\cal I}_1}-\frac{A^\mp{\cal I}_2}
{A^\pm{\cal I}_1}-\frac{A^\mp}{1+2dt}\right)K_1^\pm. 
\end{eqnarray}
We now have the complete solution of our problem, since the 
ground state (bound)  wave function is given by (13) with the
coefficients defined  in (17). In particular we find that, at $d$=1,
\begin{eqnarray}
M_2 &=&\frac{1}{4\eps^3}(h_1+h_2+\eps(l_1+l_2)),\quad
M_1=\frac{1}{4\eps}(l_1-l_2),\quad \nonumber \\
M_0& =& \frac{1}{4\eps}(\eps(l_1+l_2)-h_1-h_2),
\end{eqnarray}
with 
\begin{eqnarray}
h_{1,2} & = &\frac{1+2t}{2t}(\eps_{1,2}^2-1)^{-1/2}, \nonumber \\
l_{1,2}& = & \frac{(1+2t)^2}{4t^2}\eps_{1,2}(\eps_{1,2}^2-1)^{-3/2}.
\end{eqnarray}
After rearrangement of all the quantities we can now have a look of
our result. The 1-$d$ wave functions $\Psi_+(x)$ and   $\Psi_-(x)$ 
are plotted
in Fig.1 for a given set of parameters $\{u,v,t\}$  in the localized phase. 
In order to check out the validity of our solution, we have performed 
 some numerical simulations, and  for any set of parameters we have found
perfect agreement  with the above analytical solution.

The information of the alternating potential at the origin propagates
in the transversal direction with finite velocity, in a soliton-like fashion.
This means that if  we  look at the shape of our solution in the 
space-time manifold  we expect to find that the oscillations due to
the alternating perturbation of the potential decrease by increasing the
distance $|x|$ from  the origin and  the wave front has the shape 
of a wedge (see Fig.2).

Performing the limit $\eps \rightarrow 1^+$  we attain an unbound
state: the peaks at the origin disappear and the two wave functions
merge one over the other at diverging transversal correlation
length (i.e. at vanishing $\mu=1/\xi_\perp$).
If one  is  interested  at the behavior of the ``gap'' $\Delta_0=
\Psi_+(0)-\Psi_-(0)$
near the transition to an unbound state,  then all information
comes from the exponent $\alpha$, defined as  $\Delta_0 \simeq 
(\eps-1)^\alpha$
for $\eps\rightarrow 1^+$.
From (13) and (17) we simply find that  for $(\eps -1)\ll 1$ the gap  reads
$\Delta_0 \simeq \sqrt{(1+2t)/4}(A^+K_1^--A^-K_1^+)$ and then two 
situations are possible: (i) we can converge to the critical point for 
generic $A^+ \ne A^-$ or (ii) we can attain a delocalized phase 
for $A^+=A^-$.  More precisely, in the second hypothesis,
  we approach the unbound  state on the manifold $u=v$ and 
$u\rightarrow 0$ linearly with $\eps-1$. 
This difference is not trivial but rather 
reflects a physical property of our system: at $d=1$ it  
has been  proved that
in the two above cases the phase transition is of 
2nd. and 4th. order, respectively \cite{nz}.
By taking the dominant contributions of the integrals defined in  (9), (12)
and (18)  for $\eps\rightarrow 1^+$  in the formulas (17) for $K_1^\pm$,
after some calculations we find  that $K_1^\pm \simeq (\eps-1)^{1/4}$.
This result is  the same in both cases described above, since the limit
$u=v \rightarrow 0$ does not  modify the leading order contribution in
(17).  Then we  finally obtain that, depending  on the 
way we approach the critical state,  $\alpha=1/4$ (i)  or 
$\alpha=5/4$ (ii) (recall that $A^\pm$ linearly
converge to 0  for $\eps\rightarrow 1^+$  in the last case). 
We finally remark that our solution, in a pure statistical 
mechanics context,
has  a certain relevance because it is the partition function 
associated to Hamiltonian (2), which describes, as above explained, the
energy of a directed polymer  in a periodic potential, a problem 
with several  applications in statistical physics \cite{kpz},\cite{gg95}.  
 
In conclusion we have solved an imaginary time Schr\"odinger equation
with a non-trivial time dependent potential on a lattice.
The critical behavior separating localized and delocalized  phases has
interesting properties characterized by non trivial scaling exponents.
We expect that the counterpart  in continuum would have qualitatively
 same behavior, which may have in turn  wider implications in the 
non-equilibrium  quantum physics.

\newpage
\section*{Figure captions}

{\bf Fig. 1}\newline
The wave functions $\Psi(x)=\Psi_+(x)$ and $\Phi(x)=\Psi_-(x)$ versus
$x$ for the 1-$d$ case in the bound state.  
The curves are obtained by  smooth interpolation among the  points
given by the  analytical solution  on the lattice. Numerical simulations 
 fit exactly  the above  curves for any set $\{u,v,J\}$ of free parameters.

\vskip .7cm
{\bf Fig. 2}\newline
The full space-time shape of our wave function $\Psi(x,t)$.
 As explained in the text, the  wave front is wedge-shaped. 
The two  functions $\Psi_+$ (resp. $\Psi_-$) are easily found by 
intersecting the surface with a $t=const.$ plane   
for times $t$  such that $\Psi(x=0,t)$ reaches a  peak (resp. a valley).


\newpage

\begin{thebibliography}{99}
\bibitem {fln91} G. Forgacs, R. Lipowsky and Th.M. Nieuwenhuizen, 
in  {\it Phase Transitions and Critical Phenomena}, Vol. 14,
C. Domb and J.L. Lebowitz, eds. (Academic Press, 1991).
\bibitem{kpz} T. Halpin-Healy and Y.-C. Zhang, 
Phys. Rep. {\bf 254}, 217 (1995).
889 (1986); T.J. Newman and H. Kallabis, J. Phys. I {\bf 6}, 373 (1996);
 S. Mukherji,  Phys. Rev. E {\bf 50}, R2407 (1994).
\bibitem{azbel} M.Ya. Azbel', Phys. Rev. Lett. {\bf 73}, 138 (1994).
 \bibitem{ord} G.N. Ord, J. Phys. A {\bf 29}, L123 (1996).
\bibitem {nz}S. Nechaev and Y.-C. Zhang, Phys. Rev. Lett. {\bf 74}, 1815
(1995). (In this work only criteria for solutions are found and the 
depinning phase transition is discussed).
\bibitem{gg95} S. Galluccio and  R. Graber, Phys. Rev. E (Rapid Comm.), 
to be published (June issue).
\bibitem{foot1} Due to the symmetry respect to the origin, we   introduce
a $d$-dimensional symmetric cosine transform.
\bibitem {yd} C. Itzykson and J.-M. Drouffe, {\it Statistical Field
Theory} Vol. 1,  (Cambridge University Press, 1989). 

\end{thebibliography}
\end{document}